\documentclass{article}

    \usepackage[final,nonatbib]{neurips_mlsb_2022}

\usepackage[utf8]{inputenc} 
\usepackage[T1]{fontenc}    
\usepackage{hyperref}       
\usepackage{url}            
\usepackage{booktabs}       
\usepackage{amsfonts}       
\usepackage{nicefrac}       
\usepackage{microtype}      
\usepackage{xcolor}         
\usepackage{float}
\usepackage{bbm}
\usepackage{tabularx}
\usepackage{lineno}
\usepackage{changepage}
\usepackage{adjustbox}
\usepackage{algorithm}
\usepackage{algpseudocode}
\usepackage{amsmath,amsthm,amssymb,amsfonts, fancyhdr, color, comment, graphicx, environ}
\usepackage{todonotes}

\usepackage{amsmath}

\title{RL Boltzmann Generators for Conformer Generation in Data-Sparse Environments}

\author{%
  Yash Patel \\
  Department of Statistics\\
  University of Michigan\\
  Ann Arbor, MI 48104 \\
  \texttt{yppatel@umich.edu} \\
   \And
  Ambuj Tewari \\
  Department of Statistics\\
  University of Michigan\\
  Ann Arbor, MI 48104 \\
  \texttt{tewaria@umich.edu} \\
}

\begin{document}

\maketitle

\begin{abstract}
  The generation of conformers has been a long-standing interest to structural chemists and biologists alike. A subset of proteins known as intrinsically disordered proteins (IDPs) fail to exhibit a fixed structure and, therefore, must also be studied in this light of conformer generation. Unlike in the small molecule setting, ground truth data are sparse in the IDP setting, undermining many existing conformer generation methods that rely on such data for training. Boltzmann generators, trained solely on the energy function, serve as an alternative but display a mode collapse that similarly preclude their direct application to IDPs. We investigate the potential of training an RL Boltzmann generator against a closely related ``Gibbs score,'' and demonstrate that conformer coverage does not track well with such training. This suggests that the inadequacy of solely training against the energy is independent of the modeling modality. 
\end{abstract}

\section{Introduction}
Protein structure prediction has been a long-standing interest for physical chemists and structural biologists alike, with implications for drug discovery and environmental biopolymers. With recent advancements from AlphaFold2, previously insurmountable challenges have emerged into the realm of possibility \cite{baek2021accurate,jumper2021highly,cramer2021alphafold2}, highlighted by the introduction of new categories in the latest CASP competition, regarded as the world-stage competition in structural biology \cite{necci2021critical}. One such category is the modeling of ``protein conformational ensembles.'' AlphaFold2 has achieved unparalleled accuracy for proteins that take on fixed structures; however, many proteins of the human proteome take on multiple shapes, or ``conformations.'' Such ``intrinsically disordered proteins'' (IDPs) are not captured by current state-of-the-art protein folding solutions \cite{hawkins2017conformation,oldfield2014intrinsically,dunker2001intrinsically}. 

Recent work has investigated the use of machine learning for such conformer generation, albeit for relatively small molecules, largely inspired by techniques from generative modeling. Towards that end, works in molecular conformation generation have explored the use of variational autoencoders \cite{gupta2022artificial, minkai2021bilevel}, normalizing flows \cite{elman2019geom}, and diffusion models \cite{hoogeboom2022equivariant,xu2022geodiff,jing2022torsional}. Other works have explored more intricate, physically-inspired modeling \cite{ganea2021geomol}. Despite performing well on small molecules, specifically on the GEOM dataset \cite{axelrod2022geom}, these models all suffer from the drawback that they \textit{require} substantial training data, which ultimately come from MD simulations  or experiments. That is, to train their respective models, these approaches all required feeding in draws from the ``true'' conformer ensembles for a subset of the dataset. This, however, makes such approaches fundamentally unfit for the IDP setting, where MD simulations are completely intractable and traditional experimental methods, such as X-ray crystallography, NMR, and SAXS, fail \cite{lazar2021ped}. Cryo-EM has 
recently emerged as a potentially viable option, with recent work investigating techniques inspired by neural radiance fields to reconstruct proteins from single-particle cryo-EM images \cite{giraldo2021bayesian,cossio2018likelihood,punjani20213d,zhong2021cryodrgn}. Such reconstruction, however, currently fails to handle the large-scale structural dynamism characteristic of IDP conformers. Therefore, neither computational nor experimental methods exist for generating IDP conformers.

A potential work-around that has been explored in parallel is to train directly against the unnormalized Boltzmann distribution, explored by a subset of generative models known as ``Boltzmann generators'' \cite{noe2019boltzmann,kohler2021smooth,jing2022torsional}. However, training solely against the energy function has been observed to result in concentrated sampling on stable conformers, making the isolated use of the energy function sample inefficient and therefore unable to scale to IDPs \cite{noe2019boltzmann}. Such an empirical finding, however, has only been replicated in the context of Boltzmann generators, all of which employ normalizing flows. It is, therefore, difficult to disentangle whether such mode collapse is a function of the modeling technique or is more fundamentally linked to the isolated training against the energy. 


Another approach to avoid the direct use of the ground truth ensemble is to use reinforcement learning, as was done for simple molecules in \cite{gogineni2020torsionnet}. This approach revolves around training an agent (herein referred to as an ``RL Boltzmann generator'') with a bespoke ``Gibbs score'' reward function, which closely parallels training directly against the energy function.  This initial work, however, failed to investigate the interplay between training the RL Boltzmann generator against the Gibbs score and its final conformer coverage. Our main contribution is to demonstrate that RL Boltzmann generators solely training on the Gibbs score also exhibit symptoms of mode collapse, suggesting that this issue is more fundamentally linked to the isolated use of the energy function than to the modeling modality. Accompanying code is available at \url{https://github.com/yashpatel5400/clean_idp_rl}.

\section{Statistical Framework}
We first summarize how this problem is posited as a reinforcement learning   problem, as was originally formulated in \cite{gogineni2020torsionnet}. The ability to learn the true conformer ensemble without training data is feasible because we know the probability distribution of said ensemble follows the Boltzmann distribution,



\begin{equation}
    p(x) = \frac{1}{Z} e^{-U(x) / kT}
\end{equation}

where $Z$ is a normalizing constant, $k$ the Boltzmann constant, $T$ the temperature, and $U : \mathbb{R}^n \rightarrow \mathbb{R}$ the energy of the ``state'' of the system. For the following discussion, we fix a molecule of interest. States $\mathcal{S}$ and actions $\mathcal{A}$ are characterized by the dihedral angles of the molecule backbone. That is, $s,a \in [0,2\pi]^N$ where $N$ is the number of dihedral angles in the molecule. Actions are further discretized to be multiples of $2 \pi/M$ with $M = 6$ to improve training performance. 

Dynamics through the environment evolve through sampling of the action space from a policy $a_t\sim\pi(s_{t-1})$. A temporary state $\widehat{s_t}$ is produced by directly acting upon state $s_{t-1}$ with the torsion angles $a_{t}$. This $\widehat{s_t}$ is then relaxed using a force field $\mathcal{F}$, specifically MMFF \cite{halgren1996merck}, to produce the next state $s_t$. This procedure is then repeated $T$ times, ultimately producing a sequence of conformers $s := \{s_0, s_1, s_2, ..., s_{T}\}$. After training the RL Boltzmann generator, we expect this set of conformers $s$ to adequately cover the conformer landscape. Note that the $T$ horizon will vary from molecule to molecule depending on the expected number of conformers. 

To reward the agent, a ``Gibbs Score'' is initially formulated as follows, with $U(x)$ computed using the same classical force field used for relaxation, namely MMFF:

\begin{equation}
    \text{Gibbs}(s_t) = \frac{1}{Z_0} \exp\{-(U(s_t) - U_0) / k\tau\},
\end{equation}

where $Z_0$ and $U_0$ are normalization factors that are empirically estimated at the beginning of simulations. To encourage conformer diversity, however, the reward given is not simply the Gibbs score for the current state, but rather it after undergoing a pruning operation. If the current conformer is similar to one that has been previously observed, we do not wish to ``double reward'' the agent, else this would directly result in mode collapse. To combat this, we prune similar states before computing the reward. Denoting the current ``history'' of configurations the molecule has been in as $s := \{s_i\}_{i=1}^{t-1}$, the current conformer is pruned if

\begin{equation} 
\exists \text{ $k$ s.t. } D_{\text{TFD}}(s_t, s_k) \le \epsilon,
\end{equation}

for some pruning threshold $\epsilon$. Here, we use the Torsion Fingerprint Distance (TFD) \cite{schulz2012tfd} to compute the similarity of two conformers, although another distance metric such as RMSD may equivalently be employed. The final reward, therefore, is

\begin{equation} 
r(s_t) = \text{Gibbs}(s_t)
 \mathbbm{1}[\nexists \text{ $k$ s.t. } D_{\text{TFD}}(s_t, s_k) \le \epsilon]. 
\end{equation}


\section{Training Procedure}
To solve this learning RL problem, we employ the techniques described in \cite{gogineni2020torsionnet}, which we highlight here. The overall policy $\pi$ consists of a Graph Neural Network, which learns a set of embeddings for the molecule, followed by pooling and fully connected layers to map such an embedding to a probability vector over the action space $\mathcal{A}$. $\pi$ is then trained iteratively using PPO \cite{schulman2015trust,schulman2017proximal}.

In greater detail, the employed GNN is an edge-network MPNN \cite{fey2019fast}, which learns a set of embeddings through iterative updating using

\begin{equation} 
x_i^{t+1} = \Theta x_i^t + \sum_{j \in \mathcal{N}(i)} h(x_j^t, e_{i,j}),
\end{equation}

where $\{x_i\}_{i=1}^N$ are the embeddings for the $N$ atoms in the molecule, $e_{i,j}$ is the edge information between atoms $(i,j)$, $\Theta$ is a GRU, and $h$ is an MLP.  We initialize the atom embeddings as encodings of their locations and types.

After each iteration $t$, the output embeddings are aggregated using a set-to-set pooling operation \cite{gilmer2017neural}, which gives us an aggregated embedding of the overall molecule $y^t$. The above series of iterative transformations therefore means we end up with a history of such molecular embeddings $\{y^i\}_{i=1}^t$ at time step $t$. Since the goal is for this conformer generator to explore the landscape, taking into account previously visited conformations is critical in assessing future actions. Therefore, these $\{y^i\}$ are further passed through an LSTM to give a current ``aggregated'' state $g^t$.

To finally produce the action $a_t$, we hold fixed the list of torsion tuples that are defined for the molecule of interest, namely tuples $T_i = (b_1^i, b_2^i, b_3^i, b_4^i)$, representing the four atoms involved in the torsion. For each torsion, we pass the $g^t$ and the corresponding embeddings for the atoms through a final network, namely $p_i^t := f(x_{b_1^i}, x_{b_2^i}, x_{b_3^i}, x_{b_4^i}, g^t)$. This produces a probability vector over the action space of a \textit{single} torsion. Sampling $a_i^t\sim p_i^t$ gives the action for the \textit{single} torsion $T_i$. Repeating this for each of the torsions gives the complete action $a_t$.


To learn the dynamics, \cite{gogineni2020torsionnet} found that curricula had to be employed for complicated molecules \cite{bengio2009curriculum,wang2021survey}, since the reward landscape encountered was too sparse. We restrict ourselves to simple molecules here and, therefore, avoid the need for such curriculum learning, but further work may wish to investigate the robustness of the mode collapse found here in light of such curriculum learning.

\section{Results}
We present results of the performance of RL agent on the traditional small-molecule GEOM-Drugs dataset, following the conventions set forth in \cite{ganea2021geomol,shi2021learning}. 
Experiments with GEOM-Drugs were conducted on a randomly sampled subset of 100 molecules, using the following
evaluation metric: 

\begin{align}
    \text{COV-R}(S_g, S_r) = \frac{1}{|S_g|}\left|\{\mathcal{C}\in S_r \mid \text{RMSD}(\mathcal{C}, \widehat{\mathcal{C}}) \le \delta, \widehat{\mathcal{C}}\in S_g \}\right|
\end{align}

where $R$ represents the ``recall'' metric and $S_r,S_g$ represent the reference and generated conformer ensembles. $\delta$ is an arbitrary threshold used for designated two conformers as being ``equivalent'' for the purposes of assessing discovery. We follow the precedent of \cite{jing2022torsional} and report 
a graph of metrics over $\delta$. A similar metric can be defined for $P$, the precision, simply by exchanging $S_g$ and $S_r$.

We compare our results against the current state of the art methods, namely Torsional Diffusion, GeoMol, and FF-optimized runs of RDKit's ETKDG \cite{riniker2015better}, but largely wish to focus our discussion on the mode collapse phenomenon observed in training the RL model. The Torsional Diffusion and GeoMol models were assessed using the pre-trained models provided in the corresponding repositories, confirmed to not to have been trained on any of the molecules chosen in the test set used here. Implementation of the above described GNN models used for RL is in PyTorch-Geometric \cite{fey2019fast,paszke2019pytorch}, with the RL environment interface provided by OpenAI's gym \cite{brockman2016openai} and implementation of PPO by stable-baselines \cite{stable-baselines3}. The model was trained using Adam \cite{kingma2014adam} on an NVIDIA Tesla V100 GPU per molecule and completes in roughly two hours for 100,000 time steps.

\subsection{GEOM-Drugs}
\begin{figure}[H]
\centering
\makebox[\textwidth][c]{\includegraphics[scale=0.3]{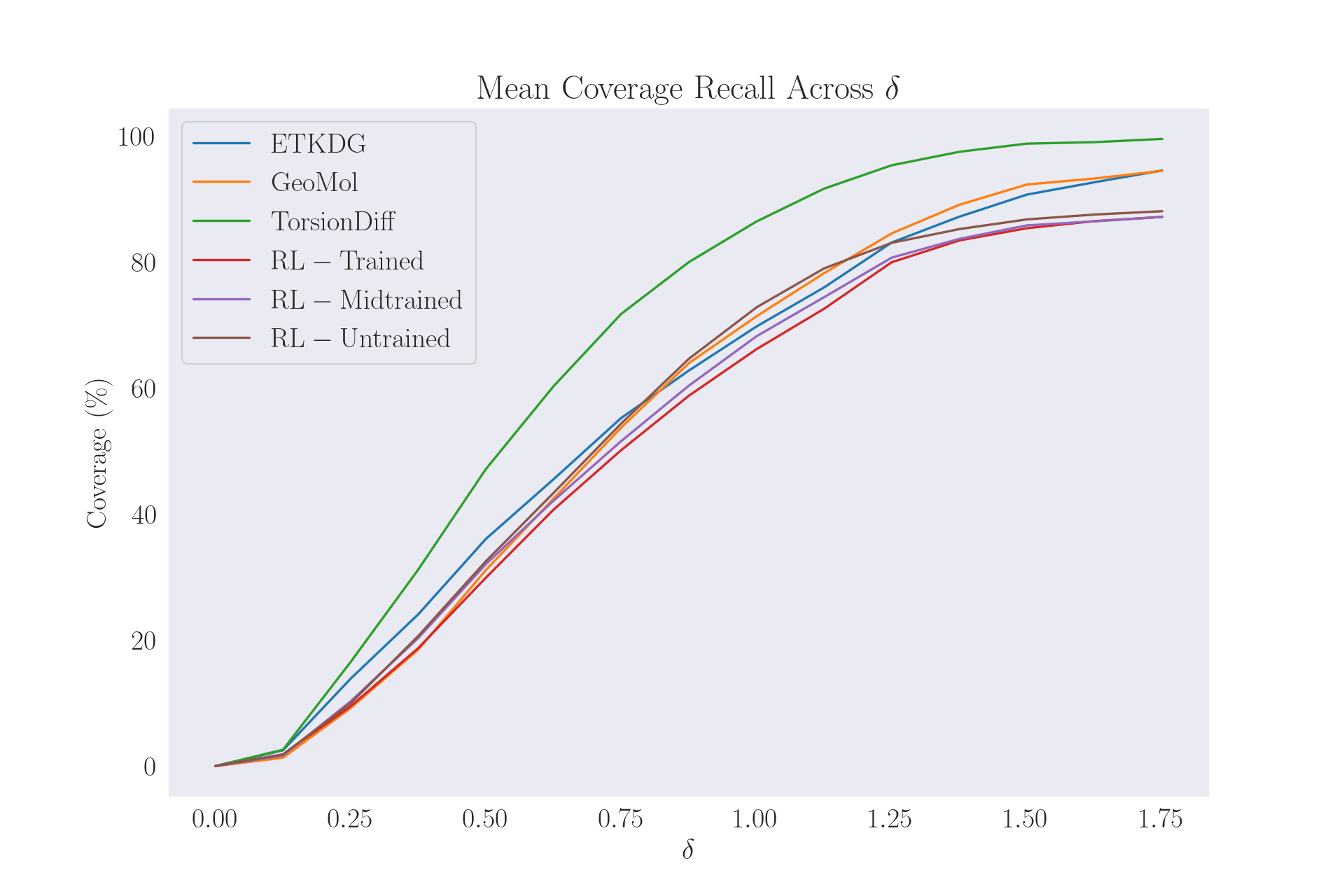}}
\caption{\label{fig:recall} Assessment of the mean coverage recalls over coverage thresholds $\delta$ for the untrained, midtrained, and trained RL agent against other state-of-the-art methods.}
\end{figure}

\begin{figure}[H]
\centering
\makebox[\textwidth][c]{\includegraphics[scale=0.3]{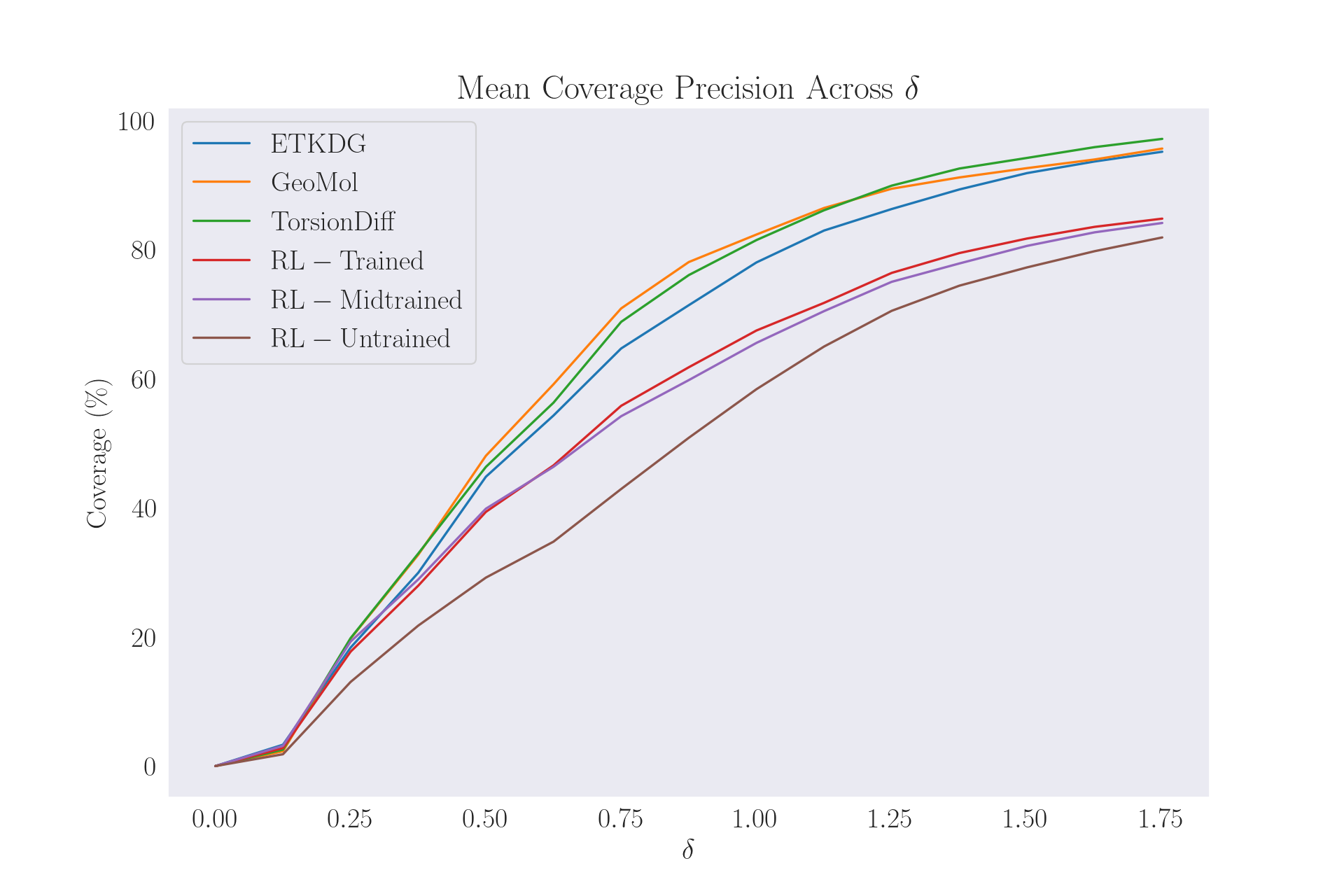}}
\caption{\label{fig:precision}  Assessment of the mean coverage precisions over coverage thresholds $\delta$ for the untrained, midtrained, and trained RL agent against other state-of-the-art methods.}
\end{figure}

We present the mean observed recalls and precisions of the RL agent on GEOM-Drugs in Figures \ref{fig:recall} and  \ref{fig:precision} respectively. The RL agent performance, for both metrics, is plotted at three stages of the training: at step 0, step 50,000, and step 100,000, respectively referred to as ``untrained,'' ``midtrained,'', and ``trained.'' We, thus, wish to concentrate this discussion on trends visible comparing these stages.

Of particular note is that the recall of the RL agent \textit{decreases} over the training run in exchange for its precision \textit{increasing}, respectively seen in the rightward shift of the recall curves and upward shifts of the precision curves for progressively more trained RL agents. This implies the RL agent conformer generation becomes more highly concentrated on a smaller subset of the conformer space, symptomatic of a mode collapse. Such a finding is particularly noteworthy given that the Gibbs score was explicitly constructed to encourage conformer diversity through pruning. An immediate question of interest, therefore, is to investigate the source of misalignment between the posited Gibbs score and coverage and whether such misalignment can be rectified.

\section{Discussion}
We have, therefore, seen that conformer generation purely using the known energy function of the Boltzmann distribution even in the form of the Gibbs score seems insufficient to obtain a diverse set of conformers. This suggests a number of directions for further investigation beyond those discussed above. A unifying theme for such direct extensions is circumventing the seeming necessity to have training data to achieve comprehensive conformer coverage while being unable to directly produce it. A current line of investigation we are pursuing is motivated by the curriculum learning employed in the previous RL Boltzmann generator work: given that MD simulations can be tractably run on small proteins or subsets of large proteins, such data can be used either to train a traditional Boltzmann generator or as a curriculum in an RL Boltzmann generator. That is, for a protein of interest, progressively longer subsequences of its amino acid chain could be taken, some of which could be feasibly analyzed with MD. For such subsequences, the Boltzmann generator could be trained with \textit{both} the energy and the generated MD ground truths, with the intention being that subsequences where no such truth is available would begin from an informed ``prior'' for energy-based sampling, resulting in enhanced conformer coverage. Pursuing this with traditional Boltzmann generators would involve a non-trivial extension to allow for such generators to be shared across different molecules, since current generators must be trained for each molecule of interest separately. 

In this vein, future work should pursue characterizing the sample efficiency of the RL generator compared to traditional Boltzmann generators, since it will point to which approach is more likely to scale to the full IDP conformer problem. That is, comparing the improvement in coverage of RL vs. traditional Boltzmann generators with each additionally provided ground truth conformer would be of great interest. Further, as briefly mentioned, the IDP problem is actively being pursued in the Cryo-EM community. With concurrent efforts in the experimental and simulation spaces, a synergy between the two is highly desirable. Integrating the method found to have the optimal sample efficiency in an iterative active learning loop with a Cryo-EM data collection pipeline would further improve the reconstruction quality and efficiency. 



\newpage
\section{Impact Statement}
As stated, this work continues in the vein of using machine learning for structural biology and, more generally, biological understanding, where given the increasing complex nature of modern scientific inquiry, it seems likely that such coordinated efforts between human ingenuity and machine learning will play an ever-increasing role in scientific exploration going forward. We expect the methods proposed here to be directly relevant for a number of biological problems, specifically in the vein of ultimately being able to study cell dynamics in silico. In the nearer term, this work will assist in understanding IDP dynamics and ultimately how these relate to their signaling functions in ligand interactions. Such work on using RL for dynamics will likely also aid in protein interaction simulations, where interpretable dynamics are of great interest. An orthogonal yet related field in which such structural dynamics are of interest is material science, in which the same methodology can be directly applied. No clear ethical concerns are posed by this work.

\bibliographystyle{unsrt}
\bibliography{references}

\section*{Checklist}

The checklist follows the references.  Please
read the checklist guidelines carefully for information on how to answer these
questions.  For each question, change the default \answerTODO{} to \answerYes{},
\answerNo{}, or \answerNA{}.  You are strongly encouraged to include a {\bf
justification to your answer}, either by referencing the appropriate section of
your paper or providing a brief inline description.  For example:
\begin{itemize}
  \item Did you include the license to the code and datasets? \answerYes{See Section.}
  \item Did you include the license to the code and datasets? \answerNo{The code and the data are proprietary.}
  \item Did you include the license to the code and datasets? \answerNA{}
\end{itemize}
Please do not modify the questions and only use the provided macros for your
answers.  Note that the Checklist section does not count towards the page
limit.  In your paper, please delete this instructions block and only keep the
Checklist section heading above along with the questions/answers below.

\begin{enumerate}

\item For all authors...
\begin{enumerate}
  \item Do the main claims made in the abstract and introduction accurately reflect the paper's contributions and scope?
    \answerYes{}
  \item Did you describe the limitations of your work?
    \answerYes{}
  \item Did you discuss any potential negative societal impacts of your work?
    \answerYes{}
  \item Have you read the ethics review guidelines and ensured that your paper conforms to them?
    \answerYes{}
\end{enumerate}

\item If you are including theoretical results...
\begin{enumerate}
  \item Did you state the full set of assumptions of all theoretical results?
    \answerNA{}
        \item Did you include complete proofs of all theoretical results?
    \answerNA{}
\end{enumerate}

\item If you ran experiments...
\begin{enumerate}
  \item Did you include the code, data, and instructions needed to reproduce the main experimental results (either in the supplemental material or as a URL)?
    \answerYes{}
  \item Did you specify all the training details (e.g., data splits, hyperparameters, how they were chosen)?
    \answerYes{}
        \item Did you report error bars (e.g., with respect to the random seed after running experiments multiple times)?
    \answerNo{}
        \item Did you include the total amount of compute and the type of resources used (e.g., type of GPUs, internal cluster, or cloud provider)?
    \answerYes{}
\end{enumerate}

\item If you are using existing assets (e.g., code, data, models) or curating/releasing new assets...
\begin{enumerate}
  \item If your work uses existing assets, did you cite the creators?
    \answerYes{}
  \item Did you mention the license of the assets?
    \answerNA{}
  \item Did you include any new assets either in the supplemental material or as a URL?
    \answerNA{}
  \item Did you discuss whether and how consent was obtained from people whose data you're using/curating?
    \answerNA{}
  \item Did you discuss whether the data you are using/curating contains personally identifiable information or offensive content?
    \answerNA{}
\end{enumerate}

\item If you used crowdsourcing or conducted research with human subjects...
\begin{enumerate}
  \item Did you include the full text of instructions given to participants and screenshots, if applicable?
    \answerNA{}
  \item Did you describe any potential participant risks, with links to Institutional Review Board (IRB) approvals, if applicable?
    \answerNA{}
  \item Did you include the estimated hourly wage paid to participants and the total amount spent on participant compensation?
    \answerNA{}
\end{enumerate}

\end{enumerate}


\end{document}